\documentclass[osajnl,twocolumn,showpacs,superscriptaddress,10pt]{revtex4-1} 

\usepackage{amsmath,amssymb,graphicx}
\usepackage{graphicx}
\usepackage[utf8]{inputenc}
\graphicspath{ {images/} }

\begin{document}

\title{Aberrated dark-field imaging systems}

\author{Mario A. Beltran}

\affiliation{Physics Department, Technical University of Denmark, DK-2800 Kgs, Lyngby, Denmark}
 
\author{David M. Paganin}
\affiliation{School of Physics and Astronomy, Monash University, Victoria 3800, Australia}

\begin{abstract}
We study generalized dark-field imaging systems. These are a subset of linear shift-invariant optical imaging systems, that exhibit arbitrary aberrations, and for which normally-incident plane-wave input yields zero output. We write down the theory for the forward problem of imaging coherent scalar optical fields using such arbitrarily-aberrated dark-field systems, and give numerical examples. The associated images may be viewed as a form of dark-field Gabor holography, utilizing arbitrary outgoing Green functions as generalized Huygens-type wavelets, with the Young-type boundary wave forming the holographic reference.     
\end{abstract}

\maketitle  

\section{Introduction} \label{Intro}

Dark-field imaging may be defined as any form of imaging in which only photons scattered from a sample reach the detection plane \cite{HechtBook}.  This implies that zero signal is registered in the absence of a scattering sample. It also implies that, for systems which record an image that bears a direct morphological relationship to the sample, the image of the sample appears embedded within a ``dark field''.  From this perspective, diffractive imaging ({\em e.g.}~crystallography \cite{SpenceBook} and coherent diffractive imaging \cite{Miao}) is a form of dark-field imaging, as is nulling interferometry \cite{NullingInterferometry1,NullingInterferometry2}, Schlieren imaging \cite{Schlieren2001} and annular dark-field imaging \cite{ADF}. 

We restrict consideration to dark-field  linear shift-invariant (LSI) imaging systems \cite{GoodmanFourierOpticsBook} utilizing coherent scalar radiation. Examples include scalar imaging systems employing monochromated light in the visible or x-ray regime, energy-filtered electron microscopy, and imaging using mono-energetic neutron beams. 

One can argue that all dark-field imaging systems are necessarily aberrated.  To see this, adopt the usual definition that the coherent transfer function, associated with an LSI system, is given by the Fourier transform of the real-space propagator (outgoing Green function) with which the input complex field is convolved to give the complex output field \cite{GoodmanFourierOpticsBook,PaganinBook}. A non-aberrated LSI imaging system is then defined as any such system for which the coherent transfer function is constant, corresponding to a real-space propagator that is a Dirac delta. This implies that any plane wave,  input into the unaberrated system, will be transformed into a plane wave exiting the non-aberrated system. Since normally-incident plane waves are blocked by dark-field systems, all LSI dark-field systems may be considered to be aberrated, insofar as their coherent transfer functions cannot be constant.   

This leads naturally to the investigation of  dark-field imaging systems for the case of arbitrary aberrations. This is the key aim of the present paper. Section~\ref{Sec:Theory} develops the underpinning theory.  Several special cases are examined in Sec.~\ref{SpecialCaseFwdProblem}, in which a single dark-field aberration is present.  We discuss some broader implications of the present work,  give some outlooks for future research, and offer some concluding remarks, in Sec.~\ref{SecDiscussion}.

\section{Theory} \label{Sec:Theory}

\begin{figure}[h]
\centering
\includegraphics[scale=0.5]{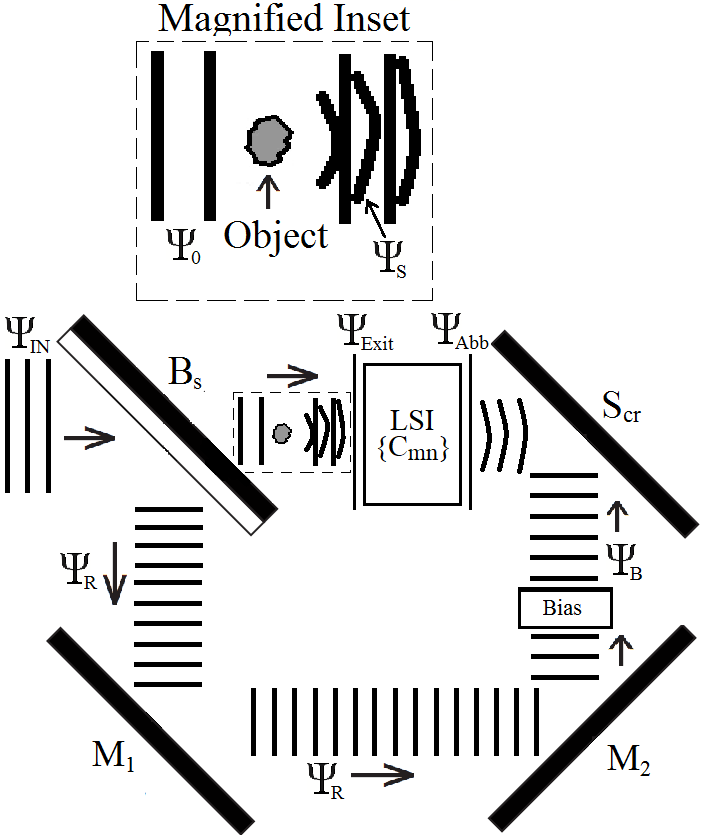}
\caption{Aberrated nulling Mach--Zehnder interferometric setup proposed to produce an aberration-induced dark-field contrast image.} 
\label{Fig:IntroDiagram}
\end{figure}

\noindent In this section we explore aberrated dark-field imaging from a classical coherent wave-optics perspective. We begin by describing the optical setup presented in Fig.~\ref{Fig:IntroDiagram}, which is a form of nulling Mach--Zehnder interferometer \cite{NullingInterferometry1,NullingInterferometry2}. Here, an incoming coherent scalar plane wave $\Psi_{\textup{IN}}$ is split into two separate plane waves $\Psi_{\textup{0}}$ (object wave) and $\Psi_{\textup{R}}$ (reference wave) by the optical beam-splitting element $\textup{B}_{\textup{s}}$. For clarity, we describe the passage of each wave-field  $\Psi_{\textup{0}}$ and $\Psi_{\textup{R}}$ separately, and eventually combine them to obtain an expression for their interference pattern on the screen.  
%
Firstly, we focus on the passage of the transmitted wave $\Psi_{\textup{0}}$. This wave-field acts as incident illumination upon the object where it undergoes phase and amplitude changes to yield an output complex wave-field at the exit surface, denoted $\Psi_{\textup{Exit}}$. This exit-surface disturbance  $\Psi_{\textup{Exit}}$ can always be decomposed as the sum of a scattered field and an unscattered field, as was done {\em e.g.}~in Gabor's development of inline holography \cite{Gabor1948}. Thus we write:
%
\begin{eqnarray} 
\Psi_{\textup{Exit}}\left ( \textbf{r} \right) =  \Psi_{\textup{S}} \left ( \textbf{r} \right)+ \Psi_{\textup{0}}\left ( \textbf{r} \right),
\label{Gabor}
\end{eqnarray}
%
\noindent where $\Psi_{\textup{S}}$ is the scattered field, $\Psi_{\textup{0}}$ is the unscattered field, and ${\bf r}=(x,y)$ denotes transverse Cartesian coordinates. The wave-field $\Psi_{\textup{Exit}}$ then traverses an aberrated linear shift-invariant imaging system. For such systems the aberrated output wave-field $\Psi_{\textup{Abb}}$, under the assumption of elastic forward scattering, is given by \cite{Allen2001,AbbBalancingPaganinGureyev,Beltran2015,Paganin2018}:  
%
\begin{eqnarray} 
\Psi_{\textup{Abb}} \left ( \textbf{r} \right)&=& \frac{1}{2\pi}\iint_{-\infty}^{\infty}d\textbf{k}_{\textbf{r}}\tilde{\Psi}_{\textup{Exit}}( \textbf{k}_{\textbf{r}}) \nonumber\\
 && \times \exp \left [ i\sum_{m,n}C_{mn}k^{m}_{x}k^{n}_{y}+i\textbf{k}_{\textbf{r}}\cdot \textbf{r}\right ]. 
\label{AberratedField}
\end{eqnarray}
%
\noindent Here, the summation is over pairs of non-negative integers $(m,n)$,  $\tilde{\Psi}_{\textup{Exit}}$ is the Fourier transform of $\Psi_{\textup{Exit}}$ with respect to ${\bf r}=(x,y)$ using the convention
%
\begin{subequations}
\begin{align}
\tilde{\Psi}( \textbf{k}_{\textbf{r}}) =\frac{1}{2\pi}\iint_{-\infty}^{\infty}d\textbf{r}\exp(-i\textbf{k}_{\textbf{r}}\cdot \textbf{r})\Psi(\textbf{r}) \label{FFTFwd}, \\
 \Psi(\textbf{r}) =\frac{1}{2\pi}\iint_{-\infty}^{\infty}d\textbf{k}_{\textbf{r}}\exp(i\textbf{k}_{\textbf{r}}\cdot \textbf{r})\tilde{\Psi}( \textbf{k}_{\textbf{r}}),
 \label{FFTInv}
\end{align}
\end{subequations}
%
\noindent $\textbf{k}_{\textbf{r}}=(k_x,k_y)$ denotes the Fourier (spatial frequency) coordinates dual to $(x,y)$, and $k = 2 \pi/\lambda$ is the wave number corresponding to the wavelength $\lambda$. The aberration coefficients of the LSI imaging system are denoted by $C_{mn}$. 

The coefficients $C_{mn}$ completely characterize the state of the aberrated imaging system. Each $C_{mn}$ coefficient is complex and can be written as \cite{AbbBalancingPaganinGureyev,Paganin2018}
%
\begin{eqnarray} 
C_{mn}=C^{(R)}_{mn} + iC^{(R)}_{mn}, 
\label{CoefficientsDefined}
\end{eqnarray}
%
\noindent where $C^{(R)}_{mn}$ and $C^{(I)}_{mn}$ are real numbers that denote the real and imaginary parts of $C_{mn}$, respectively. The set $C^{(R)}_{mn}$ denotes the coherent aberrations, with $C^{(I)}_{mn}$ being the incoherent aberrations. The connection between the complex aberration coefficients $C_{mn}$, and the Seidel aberrations of classic aberration theory, has been detailed elsewhere and will not be repeated here \cite{AbbBalancingPaganinGureyev,Paganin2018}. Note also that the coefficients $C_{mn}$ are closely related to the Cartesian representation of the Zernike polynomials, which are often used to  model the aberrations of optical imaging systems \cite{Lakshminarayanan2011}.

The aberrated wave-field $\Psi_{\textup{Abb}}$ can be obtained in terms of the scattered and unscattered wave-field  by substituting Eq.~\eqref{Gabor} into Eq.~\eqref{AberratedField}, giving
%
\begin{eqnarray} 
\Psi_{\textup{Abb}}\left ( \textbf{r} \right) =  \Psi^{\textup{Abb}}_{\textup{S}} \left ( \textbf{r} \right)+ \Psi_{0}\left ( \textbf{r} \right),
\end{eqnarray}
%
\noindent where $\Psi^{\textup{Abb}}_{\textup{S}}$ is the aberrated scattered wave-field.  Note that the normally incident plane wave $\Psi_{0}$ is completely unaffected (up to a non-zero multiplicative constant factor which is here taken to be unity) as it traverses a linear aberrated shift-invariant imaging system \cite{GoodmanFourierOpticsBook}. Consequently, only the scattered wave-field will be responsible for any disturbances in the aberrated image. We have assumed here, as we do throughout the paper, that the transfer function for the LSI imaging system does not vanish for $(k_x,k_y)=(0,0)$. The aberrated scattered field $\Psi^{\textup{Abb}}_{\textup{S}}$ is given by \cite{Allen2001,AbbBalancingPaganinGureyev}:
%
\begin{eqnarray} 
\Psi^{\textup{Abb}}_{\textup{S}} \left ( \textbf{r} \right) &=&\frac{1}{2\pi}\iint_{-\infty}^{\infty}d\textbf{k}_{\textbf{r}}\tilde{\Psi}_{\textup{S}}( \textbf{k}_{\textbf{r}}) \nonumber\\
& &\times \exp \left [ i\sum_{m,n}C_{mn}k^{m}_{x}k^{n}_{y}+i\textbf{k}_{\textbf{r}}\cdot \textbf{r}\right ]. 
\label{ScatteredAberratedField}
\end{eqnarray}
%
\noindent Here, $\tilde{\Psi}_{\textup{S}}$ denotes the Fourier transform of $\Psi_{\textup{S}}$ with respect to $x$ and $y$.

We now consider the reflected wave-field $\Psi_{\textup{R}}$, in Fig.~\ref{Fig:IntroDiagram}. Here, the mirrors $\textup{M}_{1}$ and $\textup{M}_{2}$ perform the task of re-directing $\Psi_{\textup{R}}$ towards the screen $\textup{S}_{\textup{cr}}$. However, before $\Psi_{\textup{R}}$ reaches the screen it traverses a variable phase bias device, so that $\Psi_{\textup{R}}$ experiences a constant phase shift given by
%
\begin{eqnarray} 
\Psi_{\textup{B}}({\bf r})=\Psi_{\textup{R}}({\bf r}) \exp(i\Phi_{\textup{B}}),
\label{PhaseBias}
\end{eqnarray}
%
\noindent where $\Phi_{\textup{B}}$ is a real constant.  Note that from here onwards the $\textbf{r}$ dependence of the wave-fields will dropped for simplicity.

Now that we have expressions for how both the transmitted and reflected wave-fields propagate through their respective paths toward the screen, we can write the  wave-field produced at the screen $\Psi_{\textup{Sc}}$ as  
%
\begin{eqnarray} 
\Psi_{\textup{Sc}} =  \Psi^{\textup{Abb}}_{\textup{S}}+ \Psi_{0}+ \Psi_{\textup{B}} .
\label{ScreenWaveField}
\end{eqnarray}

We now briefly consider three limit cases of Eq.~\eqref{ScreenWaveField}, corresponding respectively to (i) aberration-free bright-field imaging, (ii) aberration-free dark-field imaging and (iii) aberrated dark-field imaging.  

{\em Case \#1: Aberration-free bright-field imaging.}   Suppose that the reflected wave $\Psi_{\textup{R}}$ is completely blocked such that only the path containing the transmitted wave reaches the screen. Assume further that the imaging system has zero aberrations present ({\em i.e.},~$C_{mn}=0$ for all $m$ and $n$), so that the output field is equal to the input field ({\em i.e.,} $\Psi_{\textup{Abb}}=\Psi_{\textup{Exit}}$). Under these conditions Eq.~\eqref{ScreenWaveField} reduces to 
%
\begin{eqnarray} 
\Psi^{\textup{Bright}}_{\textup{Sc}} =  \Psi_{\textup{S}}+ \Psi_{0}.
\label{ScreenTransmittedBright}
\end{eqnarray}
%
\noindent Here, the image at the screen will detect signals due both to scattered (non-aberrated) and unscattered radiation. Such ``bright-field'' images display standard transmission contrast. For mono-energetic scalar illumination in the thin-object regime this can be calculated as:
%
\begin{eqnarray} 
\Psi^{\textup{Bright}}_{\textup{Sc}}={\mathcal{T}}\Psi_{0},
\label{ScreenTransmittedBright2}
\end{eqnarray}
%
\noindent where $\mathcal{T}$ is the complex transmission function specific to the object. The intensity is obtained by taking the modulus squared $| \Psi^{\textup{Bright}}_{\textup{Sc}}|^{2}= | \mathcal{T}\Psi_{0}  |^{2}=|{\mathcal T}|^2$, with the last equality following from the assumption that $\Psi_{0}$ is a unit-modulus plane wave. This bright-field image $|{\mathcal T}|^2$ gives a direct representation of the squared modulus of the transmission function of the thin object, and is insensitive to the phase of ${\mathcal T}$. 

{\em Case \#2: Aberration-free dark-field imaging.} Now consider a similar scenario, in which the reflected wave $\Psi_{\textup{R}}$ is unblocked, and the phase bias device is adjusted so that $\Psi_{\textup{R}}$ has been phase shifted such that when it reaches the screen we have $\Psi_{\textup{B}}=\Psi_{0} \exp(i\pi)$.  If we again assume all aberrations to vanish, Eq.~\eqref{ScreenWaveField} becomes 
%
\begin{eqnarray} 
\Psi^{\textup{Dark}}_{\textup{Sc}} &=&  \Psi_{\textup{S}} + \Psi_{0}-\Psi_{0} \nonumber\\
 &=&  \Psi_{\textup{S}}.
\label{ScatteredOnlySingal}
\end{eqnarray}
%
\noindent The unscattered and bias field now completely destructively interfere, that is $\Psi_{\textup{B}} =-\Psi_{0}$, hence we have a situation where the screen image only displays the pure scattering signal ({\em i.e.}, $ \left |\Psi^{\textup{Dark}}_{\textup{Sc}}  \right |^{2} =\left |\Psi_{\textup{S}} \right |^{2}$). 
According to Eq.~\eqref{ScatteredOnlySingal} we can calculate such a dark-field image by subtracting (via destructive interference) the unscattered wave-field $\Psi_{0}$ from the bright-field wave-field to produce an image containing only scattering information from the object:
%
\begin{eqnarray} 
\Psi^{\textup{Dark}}_{\textup{Sc}}&=&\Psi^{\textup{Bright}}_{\textup{Sc}}-\Psi_{0} \nonumber\\
&=&\mathcal{T}\Psi_{0}-\Psi_{0}.
\label{DarkInterfere}
\end{eqnarray}
%
This dark-field image exhibits phase contrast, unlike the corresponding bright-field image considered earlier (Case \#1).  For example, if the transmission function may be written as ${\mathcal T}=\exp(i\varphi)$, corresponding to a thin phase object with position-dependent phase shift $\varphi$, the corresponding nulling-interferometer intensity output $I_{\varphi}$ is the phase-dependent function
 %
\begin{eqnarray} 
I_{\varphi}=|\exp(i\varphi)-1|^2=2(1-\cos \varphi).
\label{NullingInterferogramPhaseObject}
\end{eqnarray}
%
 Note also that the above expression has the required dark-field behaviour of vanishing when $\varphi$ is zero.  
 
{\em Case \#3: Aberrated dark-field imaging.} Finally, consider the case where the aberrations in the imaging system are now non-zero ($ C_{mn}\neq 0$ for at least one pair of non-negative integers $(m,n)$). Based on Eq.~\eqref{ScreenWaveField} the wave-field at the screen will now be: 
%
\begin{eqnarray} 
\Psi^{\textup{Dark}}_{\textup{Sc}} &=&  \Psi^{\textup{Abb}}_{\textup{S}} + \Psi_{0}- \Psi_{0} \nonumber\\
 &=&  \Psi^{\textup{Abb}}_{\textup{S}}.
\label{AberratedDarkFieldImage}
\end{eqnarray}
%
\noindent It is this final scenario that constitutes aberrated dark-field imaging, as described by Eq.~\eqref{AberratedDarkFieldImage}. The forward-scattering formalism for the aberrated scattered field $\Psi^{\textup{Abb}}_{\textup{S}}$ has already been given in Eq.~\eqref{ScatteredAberratedField}. In the next section we study special cases of this equation for aberrated dark-field imaging, where the imaging systems contain a specific set of defined aberrations.

\section{Some special cases of aberrated dark-field imaging} \label{SpecialCaseFwdProblem}

\noindent Here we consider special cases of Eqs.~\eqref{AberratedField}, \eqref{ScatteredAberratedField} and \eqref{ScreenWaveField}, to exemplify the wide range of bright- and dark-field image contrast obtained from imaging systems that have a given set of aberrations. Each example is illustrated via numerical simulations. All simulated images are performed using a $512 \times 512$-pixel grid.  For simplicity, a thin single-material object is  assumed.  This implies that the projected thickness $T(i,j)$ of the object, as a function of transverse pixel coordinates $(i,j)$, can be used to calculate both the exit-surface phase map $\varphi(i,j)=k (n-1) T(i,j)$ and the exit-surface intensity $I(i,j)=\exp[-\mu T(i,j)]$ (projection approximation \cite{PaganinBook}).  Here, $n$ is the constant (real) refractive index of the thin single-material object, and $\mu$ is the corresponding linear attenuation coefficient.  

\begin{figure}[h]
\centering
\includegraphics[scale=0.69]{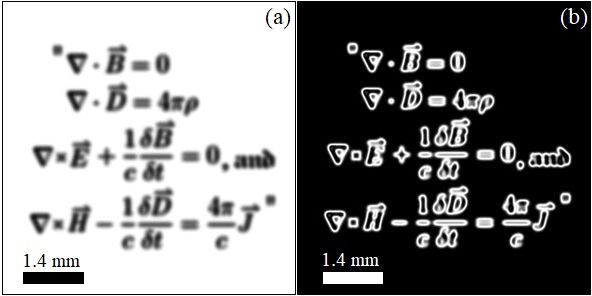}
\caption{Non-aberrated images. (a) Standard non-aberrated transmission-based bright-field image, and (b) the corresponding non-aberrated dark-field image.} 
\label{Fig:TransmissionPlane}
\end{figure}

The input projected thickness map $T(i,j)$ was taken to be a Gaussian-smoothed (to avoid edge artifacts) binary image containing the Maxwell equations, leading to the exit-surface intensity (aberration-free bright-field image) shown in Fig.~\ref{Fig:TransmissionPlane}(a).  The full-width at half maximum, of the Gaussian smoothing, was taken to be 1.5 pixels.  The linear attenuation coefficient $\mu$ was chosen so that this exit-surface intensity had a minimum intensity of 0.998 (shown as black in Fig.~\ref{Fig:TransmissionPlane}(a), corresponding to pixels $(i,j)$ where $T \ge 0$ attains its maximum value) and a maximum intensity equal to unity (shown as white in Fig.~\ref{Fig:TransmissionPlane}(a), corresponding to pixels $(i,j)$ where $T$ attained its minimum value of zero).  This corresponds to a weakly absorbing object.  The refractive index was chosen so that the associated phase shift $\varphi\equiv\arg\mathcal{T}$ varied between zero and $3.6\pi$ radians.   
%

\noindent For all the simulation results shown here the projected thickness map $T(i,j)$ is used to construct the input wave-field
\begin{eqnarray}
\nonumber \Psi_{\textup{Exit}}  &=& \left | \Psi_{\textup{Exit}} \right |e^{i\Phi_{\textup{Exit}} } \\ &=& \sqrt{\exp(-\mu T)}\exp[i k (n-1) T]
\end{eqnarray}
which is then propagated using a particular aberration via Eq.~\eqref{AberratedField}. The physical size of the image is 5.12 (W) $\times$ 5.12 (H) mm $^{2}$ and the radiation wavelength was chosen to be $\lambda=632.8$ nm, belonging to the visible light range of the electromagnetic spectrum.  


The aberration-free dark-field image may be calculated using the squared modulus of Eq.~\eqref{DarkInterfere}, as $\left | \Psi^{\textup{Dark}}_{\textup{Sc}}  \right |^{2}=\left | \mathcal{T}\Psi_{0}-\Psi_{0}  \right |^{2}$. This image is shown in Fig.~\ref{Fig:TransmissionPlane}(b).  One can clearly observe that the background is dark relative to the scattering signal.  Moreover, since the maximum phase shift is $3.6\pi$ radians and the absorption is very weak, Eq.~(\ref{NullingInterferogramPhaseObject}) implies that up to one dark band within regions of non-zero $\varphi$ is to be expected.  This is consistent with the contrast evident in Fig.~\ref{Fig:TransmissionPlane}(b).

We now turn to the main topic of this section, namely aberrated bright-field and aberrated dark-field imaging systems.  We consider defocus-aberrated bright-field and dark-field imaging in Sec.~\ref{Example1}, followed by incoherent-tilt bright-field and dark-field imaging in Sec.~\ref{Example2}.  Lastly, we treat spherical-aberration bright-field and dark-field imaging in Sec.~\ref{Example3}.

\subsubsection{Defocus-aberrated bright-field and dark-field imaging} \label{Example1}

\begin{figure}[h]
\centering
\includegraphics[scale=0.69]{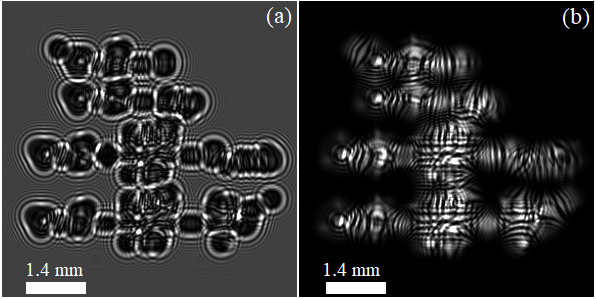}
\caption{Defocus-aberrated images. (a) Defocus-aberrated bright field image using setup in Fig.~\ref{Fig:IntroDiagram}, with $z=10.0$ mm. (b) Corresponding dark-field defocus-aberrated image. } 
\label{Fig:AberratedDefocus}
\end{figure}

\noindent For this first case we consider an aberrated system where the only non-zero coefficients present are $\left \{ C^{(R)}_{02},C^{(R)}_{20} \right \}$. The associated LSI imaging system is rotationally symmetric, corresponding to defocus (free-space propagation distance) $z$ given by \cite{PaganinBook,Allen2001,AbbBalancingPaganinGureyev,Paganin2018}:    
%
\begin{eqnarray} 
C^{(R)}_{02} &=& C^{(R)}_{20}=-\frac{z}{2k}.
\label{CoefDefocusAndBlur}
\end{eqnarray}
%
\noindent Here, $k=2\pi / \lambda$ is the radiation wave number, and we have implicitly assumed the paraxial approximation to be valid. Substitution of Eq.~\eqref{CoefDefocusAndBlur} into Eq.~\eqref{AberratedField} gives a formulation for Fresnel free-space propagation through the distance $z$: 
%
\begin{eqnarray} 
\Psi_{\textup{Abb}} = \frac{1}{2\pi}\iint_{-\infty}^{\infty}d\textbf{k}_{\textbf{r}}\tilde{\Psi}_{\textup{Exit}} \exp \left [\frac{-iz\left | \textbf{k}_{\perp } \right |^{2}}{2k} +i\textbf{k}_{\textbf{r}}\cdot \textbf{r}\right ].\nonumber\\
\label{FresnelPropagator}
\end{eqnarray}
%
For the parameters stated previously a propagated bright-field Fresnel diffraction pattern is calculated using Eq.~\eqref{FresnelPropagator}: see Fig.~\ref{Fig:AberratedDefocus}(a). Note that the reference wave-field for this example is completely blocked, so that $\Psi_{\textup{B}}=0$. The defocus distance was set to $z=10$ mm giving a Fresnel number $N_{F}=1.01$.  Since the Fresnel number is on the order of unity, the associated Fresnel diffraction pattern is in the ``intermediate field'' which lies between the near-field (where $N_{F}\gg 1$) and the far field (where $N_{F}\ll 1$) \cite{PhaseOdyssey}. Figure~\ref{Fig:AberratedDefocus}(a) shows the squared modulus of the propagated field clearly displaying Fresnel diffraction fringes resembling an inline-holographic pattern \cite{Gabor1948} in the intermediate regime, as anticipated by the order-of-unity value for the Fresnel number.  Under the formalism developed by Gabor \cite{Gabor1948}, this inline hologram may be viewed as encoding otherwise-lost phase information regarding the sample, since one has what is in essence a (phase-encoding) interferogram due to the coherent superposition of (i) the Fresnel-diffracted wave scattered by the object, and (ii) the reference wave given by the unscattered field.    
%

\noindent The associated dark-field image, shown in Fig.~\ref{Fig:AberratedDefocus}(b), is obtained by coherently superposing the propagated bright-field complex disturbance (whose intensity is given in Fig.~\ref{Fig:AberratedDefocus}(a)) with $\Psi_{\textup{B}}=-\Psi_{0}$.  This amounts to unblocking the reference field and adjusting the bias phase to $\pi$ radians. The resulting {\em dark-field hologram} encodes diffraction information solely due to the propagated scattered wave-field, implying that every photon measured at the screen has interacted with the object.  Note that our usage of the term {\em dark-field hologram} is different from an off-axis holographic crystal-strain-mapping technique of the same name (or similar name, depending on the publication), used in the electron-optics community \cite{ElectronDarkfieldHolog1,ElectronDarkfieldHolog2,Koch2010,ElectronDarkfieldHolog3,Hych2012,ElectronDarkfieldHolog4}.   

We speak of Fig.~\ref{Fig:AberratedDefocus}(b) as a dark-field {\em hologram}, because it is a form of diffraction pattern that encodes both phase and intensity information regarding a field.  This claim may be made despite the fact that the ``usual'' inline-holography reference wave, namely the unscattered wave, has been removed in the nulling-interferometry dark-field setup.  We are justified in speaking of aberrated dark-field images (such as Fig.~\ref{Fig:AberratedDefocus}(b)) as ``dark-field {\em holograms}'', since such images may be regarded in holographic--interferometric terms: they are the interference pattern resulting from the coherent superposition of (i) the wave that is transmitted through the sample, and (ii) the so-called boundary-diffraction wave that is scattered from the edge of the object.  This follows from the concept of the Young--Maggi--Rubinowicz boundary wave, first discussed qualitatively by Young \cite{YoungOnTheBoundaryWave}, with corresponding theory developed by Maggi \cite{Maggi}, Rubinowicz \cite{Rubinowicz}, and Miyamoto \& Wolf \cite{MiyamotoWolf1, MiyamotoWolf2}. If one works in an asymptotic (short-wavelength) setting using  complex rays, the same conclusion can be drawn based on Keller's geometric theory of diffraction \cite{Keller}, with the boundary wave being replaced by complex rays diffracted from the edge of the object. We consider these points further, in Sec.~\ref{SecDiscussion} below.

\subsubsection{Incoherent-tilt aberrated bright-field and dark-field imaging} \label{Example2}

\noindent The next example considers an aberrated system where the only non-zero coefficients present are incoherent.  This corresponds to all non-zero aberration coefficients $C_{mn}$ being purely imaginary numbers.  We consider incoherent tilt $\left \{ C^{(I)}_{01},C^{(I)}_{10} \right \}$ \cite{PaganinLinearPropagators}. 

Equation~\eqref{AberratedField} gives: 
%
\begin{eqnarray} 
\Psi_{\textup{Sc}}= \frac{1}{2\pi}\iint_{-\infty}^{\infty}d\textbf{k}_{\textbf{r}}\tilde{\Psi}_{\textup{Exit}} \exp \left [-A_{t} \textbf{k}_{\textbf{r}}\cdot \textbf{n}_{\textbf{k}_{\textbf{r}}} +i\textbf{k}_{\textbf{r}}\cdot \textbf{r}\right ].\nonumber\\
\label{IncohTiltPropagator}
\end{eqnarray}
\noindent Here,
\begin{equation}
A_{t}=\sqrt{(C^{(I)}_{01})^2+(C^{(I)}_{10})^2}, 
\end{equation}
which has units of length, is the incoherent-tilt
aberration coefficient, and
\begin{equation}
\textbf{n}_{\textbf{k}_{\textbf{r}}}=(\cos\alpha,\sin\alpha)
\end{equation}
is a unit vector in the $(k_x,k_y)$ plane, making an angle of $\alpha$ radians with respect to the positive-$k_x$ axis.

\begin{figure}[h]
\centering
\includegraphics[scale=0.69]{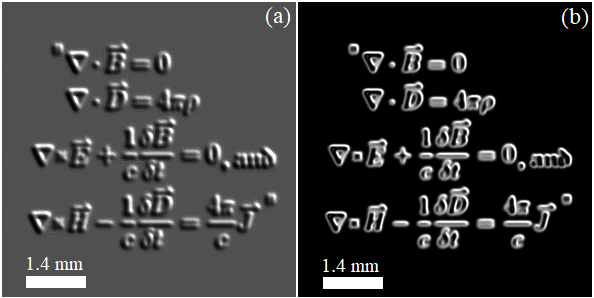}
\caption{Incoherent-tilt aberrated images. (a) Defocus-aberrated bright field image using setup in Fig.~\ref{Fig:IntroDiagram}, with $A_{t}=3.0$ $\mu$m. (b) Corresponding  dark-field incoherent-tilt aberrated image (dark-field hologram). } 
\label{Fig:IncohTilt}
\end{figure}

A propagated bright-field incoherent-tilt aberrated image was calculated using Eq.~\eqref{IncohTiltPropagator}, corresponding to $\alpha=\pi/2$ and $A_t=3.0$ $\mu$m, and is shown in Fig.~\ref{Fig:IncohTilt}(a) ($\Psi_{\textup{B}}=0$). Notice the dark-to-bright gradient across the x--y direction, resembling a pattern normally found in images that display differential phase contrast \cite{PaganinLinearPropagators}.  Examples of such contrast include differential interference contrast for visible light optics \cite{VisibleLightDIC}, together with analyzer-crystal phase contrast for x-ray optics \cite{ABI1,ABI2}. 
%

\noindent The associated dark-field image shown in Fig.~\ref{Fig:IncohTilt}(b) was computed by setting $\Psi_{\textup{B}}=-\Psi_{0}$. Here, the edges are completely preserved as in Fig.~\ref{Fig:TransmissionPlane}(b) where only transmission contrast is considered. The contrast seen here also resembles images where a Sobel operator is applied as a method of edge detection \cite{Sobel}.     

\subsubsection{Spherically-aberrated bright-field and dark-field imaging} \label{Example3}

\noindent Our final example considers an aberrated system where the only non-zero coefficients present are taken from the set  $\{C^{(R)}_{04},C^{(R)}_{40},\frac{1}{2}C^{(R)}_{22} \}$ \cite{AbbBalancingPaganinGureyev}. If we take 
%
\begin{eqnarray} 
C^{(R)}_{04}=C^{(R)}_{40}=\frac{1}{2}C^{(R)}_{22}=\frac{-C_{S}}{8k^{3}},
\label{CoefSpherical}
\end{eqnarray}
%
\noindent then, like defocus, the resulting LSI imaging system has rotational symmetry.  The coefficient $C_S$, which has units of length, denotes spherical aberration. Substituting Eq.~\eqref{CoefSpherical} into Eq.~\eqref{AberratedField} gives:
%
\begin{eqnarray} 
\Psi_{\textup{Sc}} = \frac{1}{2\pi}\iint_{-\infty}^{\infty}d\textbf{k}_{\textbf{r}}\tilde{\Psi}_{\textup{Exit}} \exp \left [\frac{-iC_{S}\left | \textbf{k}_{\perp} \right |^{4}}{8k^{3}} +i\textbf{k}_{\textbf{r}}\cdot \textbf{r}\right ]. \nonumber\\
\label{SphericalPropagator}
\end{eqnarray}

\begin{figure}[h]
\centering
\includegraphics[scale=0.69]{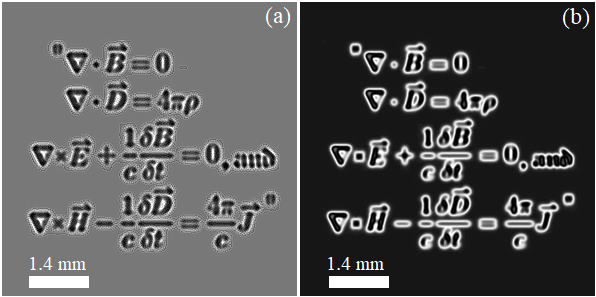}
\caption{Spherically aberrated images. (a) Spherically-aberrated bright-field image using setup in Fig.~\ref{Fig:IntroDiagram}, with $C_{S}=5.0$ mm. (b) Corresponding dark-field spherically-aberrated image (dark-field hologram).} 
\label{Fig:Spherical}
\end{figure}

Figure~\ref{Fig:Spherical}(a) shows a propagated bright-field spherically aberrated intensity image calculated using Eq.~\eqref{SphericalPropagator} ($\Psi_{\textup{B}}=0$). Here, $C_S=5.0$ mm.  Note that the bi-Laplacian phase contrast seen here is similar to that of near-field ($N_{f}\gg1$) Laplacian-contrast defocus images often seen in propagation-based x-ray phase-contrast  \cite{SnigirevPBI,CloetensPBI,WilkinsPBI} and out-of-focus electron microscopy \cite{CowleyBook}. For a specific study of bi-Laplacian phase contrast due to spherical aberration, see Lynch {\em et al.} \cite{Lynch} and Paganin \& Gureyev \cite{AbbBalancingPaganinGureyev}.  This link between (Laplacian-type) defocus phase contrast and (bi-Laplacian-type) spherical-aberration phase contrast is related to the fact that the Huygens wavelets (outgoing spherical waves) associated with the Fresnel diffraction theory (defocus aberration), have been replaced with a different but nevertheless rotationally-symmetric generalized Huygens wavelet (outgoing Green function) that is associated with pure spherical aberration.  

%

\noindent The corresponding dark-field image is shown in Fig.~\ref{Fig:Spherical}(b), computed by setting $\Psi_{\textup{B}}=-\Psi_{0}$. Again, as with the dark-field examples in Fig.~\ref{Fig:TransmissionPlane}(b) and Fig.~\ref{Fig:IncohTilt}(b), the edges are also completely preserved.  We may again view this as a form of aberrated dark-field hologram, using similar reasoning to that given earlier, but with the outgoing Green function (generalized Huygens-type wavelet) for paraxial free-space propagation being replaced by the previously-mentioned outgoing Green function corresponding to pure spherical aberration.

\section{Discussion and Summary} \label{SecDiscussion}
 
\noindent A variety of aberrated dark-field intensity images (dark-field holograms) was seen to be possible, within the framework of linear shift-invariant imaging systems. Simulation examples were provided for specific types of aberration, in both bright-field and dark-field modalities: defocus, incoherent tilt and spherical aberration. Some of the examples display unfamiliar contrast, since the associated outgoing Green function is not the expanding spherical wave that is familiar {\em e.g.}~from the Fresnel theory of diffraction. Unlike dark-field imaging systems that utilize some kind of optical filtering strategy to ``block'' the unscattered field ($\Psi_{0}$), which in practice would result in also blocking part of the scattered field ($\Psi_{\textup{S}}$), our formalism relies on pure wave interference (nulling interferometry) that completely removes $\Psi_{0}$ and avoids any such blockage of $\Psi_{\textup{S}}$. Nevertheless, our formalism (Eq.~\ref{ScatteredAberratedField}) would still hold for the former case where the unscattered wave is blocked---{\em e.g.} for aberrated imaging systems whose complex transfer function vanishes when $(k_x,k_y)=(0,0)$--- provided that this only filters a negligibly small proportion of the scattered wave-field.   

The focus of this paper has been the forward problem of aberrated dark-field imaging.  This immediately raises the corresponding inverse problem of how to recover wave-field phase information from such images. One might seek to solve this via iterative approaches {\em e.g.}~along the lines pioneered by Gerchberg and Saxton \cite{GerchbergSaxton} and Fienup \cite{Fienup}, however deterministic methods such as those based on continuity equations governing the flow of energy of optical fields have proven to be a formidable option \citep{PagNug1998,Teague1983}. Certain recently-developed transport equations, for the evolution of optical intensity as a function of evolving the aberrations associated with LSI imaging systems, might be adapted to enable deterministic avenues for the inverse problem of phase retrieval from aberrated dark-field holograms \cite{Paganin2018}.  Another interesting topic for future investigation is dark-field aberrated imaging of fields containing phase vortices, such as those found in near field speckle patterns using spatially-random phase--amplitude screens \cite{GoodmanSpeckleBook}, far-field diffraction patterns from most non-trivial scatterers \cite{PaganinBook}, and the focal volume of coherently-illuminated aberrated lenses \cite{BoivinDowWolf}. 

We close with a further exploration of the concept of an {\em aberrated dark-field hologram}.  This builds on Gabor's original conception of inline holograms as encoding information regarding the whole ({\em i.e.} both amplitude and phase) of a coherent wave-front, since such conventional inline holograms may be viewed in interferometric terms as an interference pattern (which therefore encodes both phase and amplitude information, rather than just amplitude information) \cite{Gabor1948}.  This interference pattern is between unscattered and scattered fields (in inline holography), thus eliminating the need for a separate reference wave (as {\em e.g.}~in the later development of off-axis holography by Leith and Upatnieks \cite{LeithUpatnieks1,LeithUpatnieks2,LeithUpatnieks3}).  

We extend this idea via the term {\em dark-field hologram}, applied to our aberrated dark-field images.  There are at least three related reasons justifying the term {\em hologram} in this context.  (i)  Under the boundary-wave theory of diffraction \cite{YoungOnTheBoundaryWave,Maggi,Rubinowicz,MiyamotoWolf1,MiyamotoWolf2} and the assumption of an optically thin phase--amplitude object, the aberrated dark-field image may be viewed as resulting from the interference between the complex transmitted wave field that would be predicted for the exit-surface of the object using the projection approximation, and the wave scattered from the edges of a compact object that is entirely contained within the field of view of the illuminating beam, in an aberrated dark-field imaging system. (ii) Under the geometric theory of diffraction, initiated by Keller \cite{Keller}, a very similar view holds, but with waves replaced by complex rays that are diffracted when they encounter the edges of the object and/or the edge of any sharp confining aperture.  (iii) In the short-wavelength limit, the asymptotic formulation of the diffraction integral associated with an aberrated dark-field imaging system, may be viewed as superposing contributions due to interior points in the geometric shadow of an illuminated object (critical points of the first kind), and points on the boundary of the illuminated object (critical points of the second kind) \cite{MandelWolf}.  

\begin{figure}[h]
\centering
\includegraphics[scale=0.45]{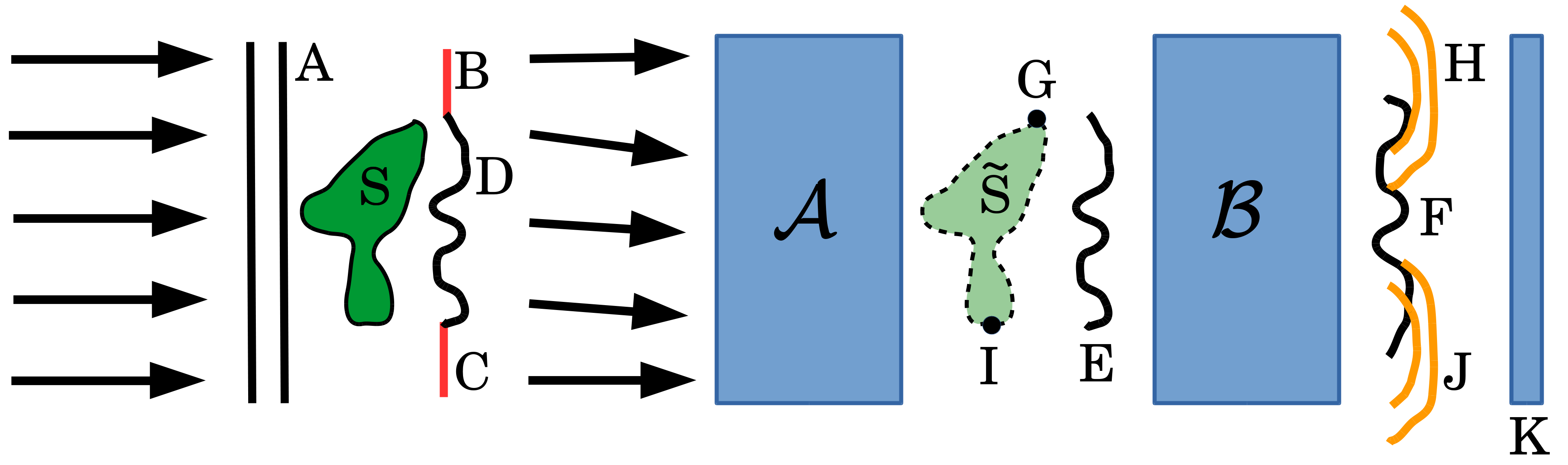}
\caption{Setup for recording aberrated dark-field holograms.} 
\label{Fig:DarkfieldHolography}
\end{figure}

\noindent The above considerations lead to the concept of dark-field holography as illustrated in Fig.~\ref{Fig:DarkfieldHolography}.  Here, a compact source of coherent scalar paraxial radiation (not shown, located to the far left of the diagram) generates planar wave-fronts $A$ that impinge upon a thin compact sample $S$.  The field at the exit surface of the sample may be decomposed into wave-fronts such as $B$ and $C$ that lie outside the geometric shadow of the sample, and wave-fronts such as $D$ that either lie within the geometric shadow of the sample, or on its boundary.  The subsequent action of the aberrated dark-field imaging system may be viewed as a two-step process. Firstly, the nulling interferometer $\mathcal{A}$ removes the background field, creating a real non-aberrated dark-field image $\tilde{S}$ of the sample which is non-zero only over the geometric-shadow region of the compact object.  The associated exit-surface dark-field disturbance $E$ then propagates through the aberrated LSI imaging system $\mathcal{B}$ to give a disturbance $\Psi_{\textup{Sc}}^{\textrm{Dark}}$ whose intensity is the registered aberrated dark-field image.  This encodes phase information regarding the wave-field at the exit-surface of the object, since it may be viewed as the interferogram over the surface of the position-sensitive detector $K$, resulting from (i) the field $F$ generated by the action of the LSI imaging system on the wave-fronts $E$, superposed with (ii) boundary-wave fields such as $H$ and $J$ resulting from the points $G$ and $I$ respectively. 

The key differences, between this scenario and that of conventional inline holography, are (i) the fact that the unscattered wave is removed, with the reference wave now being composed of the boundary wave, and (ii) the Green function (generalized Huygens wavelet) is now given by the inverse Fourier transform (with respect to $k_x$ and $k_y$) of the complex transfer function for the LSI imaging system, rather than being restricted to an expanding spherical wave. Under this view, the fringes evident in the conventional inline hologram in Fig.~\ref{Fig:AberratedDefocus}(a) are due to the interference between (i) the wave that is transmitted through the sample and (ii) the unscattered wave, while the fringes evident in Fig.~\ref{Fig:AberratedDefocus}(b) are due to the interference between (iii) the object-transmitted wave ($F$ in Fig.~\ref{Fig:DarkfieldHolography}) and (iv) its associated boundary wave ($H$ and $J$ in Fig.~\ref{Fig:DarkfieldHolography}).

\section*{Acknowledgements}

\noindent The authors received financial support from a VILLUM Experiment grant. In addition, they gratefully acknowledge helpful discussions with Hugh Simons (Physics Department, Technical University of Denmark) and Timothy Petersen (School of Physics and Astronomy, Monash University, Australia).


\begin{thebibliography}{99}

\bibitem{HechtBook} E. Hecht, {\em Optics}, 5th ed. (Pearson Education Ltd., Boston, 2017).

\bibitem{SpenceBook} J. C. H. Spence, {\em High-Resolution Electron Microscopy}, 4th ed. (Oxford University Press, Oxford, 2017).

\bibitem{Miao} J. Miao, P. Charalambous, J. Kirz, and D. Sayre, {\em Extending the methodology of X-ray crystallography to allow imaging of micrometre-sized non-crystalline specimens}, Nature {\bf 400}, 342--344 (1999).

\bibitem{NullingInterferometry1} R. N. Bracewell, {\em Detecting nonsolar planets by spinning infrared interferometer}, Nature {\bf 274}, 780--781 (1978).

\bibitem{NullingInterferometry2} P. M. Hinz,  J. R. P. Angel, W. F. Hoffmann, D. W. McCarthy Jr, P. C. McGuire, M. Cheselka, J. L. Hora, and N. J. Woolf, {\em Imaging circumstellar
environments with a nulling interferometer}, Nature {\bf 395},251--253 (1998).

\bibitem{Schlieren2001} G. S. Settles, {\em Schlieren and Shadowgraph Techniques} (Springer, Berlin, 2001).

\bibitem{ADF} M. De Graef, {\em Introduction to Conventional Transmission Electron Microscopy} (Cambridge University Press, Cambridge, 2003).

\bibitem{GoodmanFourierOpticsBook} J. W. Goodman, {\em Introduction to Fourier optics}, 3rd ed. (Roberts \& Company, Englewood Colorado, 2005).

\bibitem{PaganinBook} D. M. Paganin, {\em Coherent X-Ray Optics} (Oxford University Press, Oxford, 2006).

\bibitem{Gabor1948} D. Gabor, {\em A new microscopic principle}, Nature {\bf 161}, 777--778 (1948).

\bibitem{Allen2001} L. J. Allen, M. P. Oxley, and D. Paganin, {\em Computational aberration correction for an arbitrary linear imaging system}, Phys. Rev. Lett. {\bf 87}, 123902-4 (2001). 

\bibitem{AbbBalancingPaganinGureyev} D.M. Paganin and T.E. Gureyev, {\em Phase contrast, phase retrieval and aberration balancing in shift-invariant linear imaging systems}, Opt. Commun. {\bf 281}, 965--981 (2008). 

\bibitem{Beltran2015} M. A. Beltran, M. J. Kitchen, T. C. Petersen, and D. M. Paganin, {\em Aberrations in shift-invariant linear optical imaging systems using partially coherent fields}, Opt. Commun. {\bf 355}, 398–405 (2015).

\bibitem{Paganin2018} D. M. Paganin, T. C. Petersen, and M. A. Beltran, {\em Propagation of fully coherent and partially coherent complex scalar fields in aberration space}, Phys. Rev. A. {\bf 97}, 023835 (2018). 

\bibitem{Lakshminarayanan2011} V. Lakshminarayanan and A. Fleck, {\em Zernike polynomials: A
guide}, J. Mod. Opt. {\bf 58}, 545--561 (2011).

\bibitem{PhaseOdyssey} K. A. Nugent, D. Paganin, and T. E. Gureyev, {\em A phase odyssey}, Phys. Today {\bf 54}:8, 27--32 (2001). 

\bibitem{ElectronDarkfieldHolog1} M. H\"{y}tch, F.  Houdellier, F. H\"{u}e and E. Snoeck, {\em Nanoscale holographic interferometry for strain
measurements in electronic devices}, Nature {\bf 453}, 1086--1089 (2008).

\bibitem{ElectronDarkfieldHolog2} M. H\"{y}tch, F.  Houdellier, F. H\"{u}e and E. Snoeck, {\em Dark-field electron holography for the mapping of strain in nanostructures: correcting artefacts and aberrations}, J. Phys.: Conf. Ser. {\bf 241} 012027 (2010).

\bibitem{Koch2010} C. T. Koch, V. B. \"{O}zd\"{o}l, and P. A. van Aken, {\em An efficient, simple, and precise way to map strain with nanometer resolution in semiconductor devices}, Appl. Phys. Lett. {\bf 96}, 091901 (2010). 

\bibitem{ElectronDarkfieldHolog3} A. B\'{e}ch\'{e}, J. L. Rouvi\`{e}re, J. P. Barnes and D. Cooper, {\em Dark field electron holography for strain measurement}, Ultramicroscopy {\bf 111}, 227--238 (2011).

\bibitem{Hych2012} M. H\"{y}tch, C. Gate, F. Houdellier, E. Snoeck, and K. Ishizuka, {\em Darkfield electron holography for strain mapping at the nanoscale}, Microsc. Anal. {\bf 26}(7), 6--10 (2012). 

\bibitem{ElectronDarkfieldHolog4} T. Denneulin and M. H\"{y}tch, {\em Four-wave dark-field electron holography for imaging strain fields}, J. Phys. D: Appl. Phys. {\bf 49}, 244003 (2016).

\bibitem{YoungOnTheBoundaryWave} T. Young, {\em The Bakerian lecture: On the theory of light and colours}, Phil. Trans. R. Soc. Lond. {\bf 92}, 12--48 (1802).

\bibitem{Maggi} G. A. Maggi, {\em Sulla propagazione libera e perturbata delle onde luminose in un mezzo isotropo}, Annali di Mat. (2), {\bf 16}, 21--48 (1888).

\bibitem{Rubinowicz} A. Rubinowicz, {\em Die Beugungswelle in der Kirchhoffschen Theorie der Beugungserscheinungen}, Ann. Physik, {\bf 53}, 257--278 (1917).

\bibitem{MiyamotoWolf1} K. Miyamoto and E. Wolf, {\em Generalization of the Maggi--Rubinowicz theory of the boundary diffraction wave--Part {I}}, J. Opt. Soc. Am. {\bf 52}, 615--625 (1962).

\bibitem{MiyamotoWolf2} K. Miyamoto and E. Wolf, {\em Generalization of the Maggi--Rubinowicz theory of the boundary diffraction wave--Part {II}}, J. Opt. Soc. Am. {\bf 52}, 626--637 (1962).

\bibitem{Keller} J. B. Keller, {\em Geometrical theory of diffraction}, J. Opt. Soc. Am. {\bf 52}, 116--130 (1962).

\bibitem{PaganinLinearPropagators} D. Paganin, T. E. Gureyev, K. M. Pavlov, R. A. Lewis, and M. Kitchen, {\em Phase retrieval using coherent imaging systems with linear transfer functions}, Opt. Commun. {\bf 234}, 87--105 (2004).

\bibitem{VisibleLightDIC} G. Nomarski and A. R. Weill, {\em Application \`{a} la m\'{e}tallographie des m\'{e}thodes interf\'{e}rentielles \`{a} deux ondes polarise\'{e}s}, Rev. Metall. {\bf 2}, 121--128 (1955).

\bibitem{ABI1} E. F\"{o}rster, K. Goetz, and P. Zaumseil, {\em Double crystal diffractometry for the characterization of targets for laser fusion experiments}, Krist. Tech. {\bf 15}, 937--945 (1980).

\bibitem{ABI2} V. A. Somenkov, A. K. Tkalich, and S. Shil'stein, {\em Refraction contrast in x-ray microscopy}, Sov. Phys. Tech. Phys. {\bf 3}, 1309--1311 (1991).

\bibitem{Sobel} I. Sobel and G. Feldman, {\em A $3\times 3$ isotropic gradient operator for image processing}, unpublished talk presented at Stanford Artificial Intelligence Project (1968). 

\bibitem{SnigirevPBI} A. Snigirev, I. Snigireva, V. Kohn, S. Kuznetsov, and I. Schelokov, {\em On the possibilities of X-ray phase contrast microimaging by coherent high-energy synchrotron radiation}, Rev. Sci. Instrum. {\bf 66}, 5486--5492 (1995).

\bibitem{CloetensPBI} P. Cloetens, R. Barrett, J. Baruchel, J.-P. Guigay, and M. Schlenker, {\em Phase objects in synchrotron radiation hard X-ray imaging}, J. Phys. D: Appl. Phys. {\bf 29}, 133--146 (1996). 

\bibitem{WilkinsPBI} S.W. Wilkins, T.E. Gureyev, D. Gao, A. Pogany and A.W. Stevenson, {\em Phase-contrast imaging using polychromatic hard X-rays}, Nature {\bf 384}, 335--338 (1996). 

\bibitem{CowleyBook} J. M. Cowley, {\em Diffraction Physics}, 3rd ed. (North--Holland, Amsterdam, 1995).

\bibitem{Lynch} D. F. Lynch, A. F. Moodie and M. A. O'Keefe, {\em n-beam lattice images. V. The use of the charge-density approximation in the interpretation of lattice images}, Acta Cryst. A {\bf 31}, 300--307 (1975).

\bibitem{GerchbergSaxton} R. W. Gerchberg and W. O. Saxton, {\em A practical algorithm for the determination of phase from image and diffraction plane pictures}, Optik {\bf 35}, 237--246 (1972).

\bibitem{Fienup} J. R. Fienup, {\em Phase retrieval algorithms: a comparison}, Appl. Opt. {\bf 21}, 2758--2769 (1982).

\bibitem{PagNug1998} D. Paganin and K. A. Nugent, {\em Noninterferometric phase imaging with partially coherent light}, Phys. Rev. Lett. {\bf 80}, 2586-2589 (1998).

\bibitem{Teague1983} M. R. Teague, {\em Deterministic phase retrieval: a Green's function solution}, J. Opt. Soc. Am. {\bf 73}, 1434--1441 (1983).

\bibitem{GoodmanSpeckleBook} J. W. Goodman, {\em Speckle Phenomena in Optics} (Roberts \& Company, Greenwood Village Colorado, 2007).

\bibitem{BoivinDowWolf} A. Boivin, J. Dow, and E. Wolf, {\em Energy flow in the neighborhood of the focus of a coherent beam}, J. Opt. Soc. Am. {\bf 57}, 1171--1175 (1967).

\bibitem{LeithUpatnieks1} E. N. Leith and J. Upatnieks, {\em Reconstructed wavefronts and communication theory}, J. Opt. Soc. Am, {\bf 52}, 1123--1130 (1962). 

\bibitem{LeithUpatnieks2} E. N. Leith and J. Upatnieks, {\em Wavefront reconstruction with continuous-tone objects}, J. Opt. Soc. Am, {\bf 53}, 1377--1381 (1963).

\bibitem{LeithUpatnieks3} E. N. Leith and J. Upatnieks, {\em Wavefront reconstruction with diffused illumination and three-dimensional objects}, J. Opt. Soc. Am, {\bf 54}, 1295--1301 (1964).

\bibitem{MandelWolf} L. Mandel and E. Wolf, {\em Optical Coherence and Quantum Optics} (Cambridge University Press, Cambridge, 1995).





\end{thebibliography}
\end{document}